\documentclass{aastex}
\usepackage{spr-astr-addons}
\usepackage{url}

\RequirePackage{color}

\def\br{Br$\gamma$}

\def\feii{[Fe\,{\sc ii}]}

\def\hm{H$_2$}

\def\p1{Paper~I}

\def\kms {$\rm km\,s^{-1}$}

\begin{document}

\title{The origin of the near-IR line emission from molecular, low and high ionization gas in the inner kiloparsec of NGC\,6240}
\slugcomment{Not to appear in Nonlearned J., 45.}
\shorttitle{The origin of the near-IR line emission in the inner kiloparsec of NGC\,6240}
\shortauthors{Ilha, Bianchin \& Riffel}

\author{Gabriele da Silva Ilha}
\affil{Universidade Federal de Santa Maria, Departamento de F\'\i sica, Centro de Ci\^encias Naturais e Exatas, 
97105-900, Santa Maria, RS, Brazil}
\and
\author{Marina Bianchin}
\affil{Universidade Federal de Santa Maria, Departamento de F\'\i sica, Centro de Ci\^encias Naturais e Exatas, 
97105-900, Santa Maria, RS, Brazil}
\and
\author{Rogemar A. Riffel}
\affil{Universidade Federal de Santa Maria, Departamento de F\'\i sica, Centro de Ci\^encias Naturais e Exatas, 
97105-900, Santa Maria, RS, Brazil}


\begin{abstract}

The understating of the origin of the H$_2$ line emission from the central regions of galaxies represent an important key to improve our knowledge about the excitation and ionization conditions of the gas in these locations. Usually these lines can be produced by Starburts, shocks and/or radiation from an active galactic nucleus (AGN). Luminous Infrared Galaxies (LIRG) represent ideal and challenging objects to investigate the origin of the H$_2$ emission, as all processes above can be observed in a single object. In this work, we use K-band integral field spectroscopy to map the emission line flux distributions and kinematics and investigate the origin of the molecular and ionized gas line emission from inner 1.4$\times$2.4 kpc$^{2}$ of the LIRG NGC\,6240, known to be the galaxy with strongest H$_2$ line emission. The emission lines show complex profiles at locations between both nuclei and surrounding the northern nucleus, while at locations near the southern nucleus and at 1$^{\prime\prime}$ west of the northern nucleus, they can be reproduced by a single gaussian component. We found that the H$_2$ emission is originated mainly by thermal processes, possible being dominated by heating of the gas by X-rays from the AGN at locations near both nuclei. For the region between the northern and southern nuclei shocks due to the interacting process may be the main excitation mechanism, as indicated by the high values of the H$_2\,\lambda 2.12\,\mu$m/\br\ line ratio. A contribution of fluorescent excitation may also be important at locations near 1$^{\prime\prime}$ west of the northern nucleus, which show the lowest line ratios. The [Fe\,{\sc ii}]$\lambda2.072\,\mu$m/\br\ ratio show a similar trend as observed for H$_2\,\lambda 2.12\,\mu$m/\br, suggesting that  \feii\ and H$_2$ line emission have similar origins. Finally, the [Ca\,{\sc viii}]$\lambda2.32\mu$m coronal line emission is observed mainly in regions next to the nuclei, suggesting it is originated gas ionized by the radiation from the AGN. 
 
\end{abstract}




\section{Introduction}
\label{sec:intro}

Merging galaxies are one of the keys to understand Starbursts, nuclear activity and galaxy evolution. Most of them are classified as Luminous InfraRed Galaxies (LIRGs) or UltraLuminous InfraRed Galaxies (ULIRGs), with infrared luminosities in the ranges $L_{IR}=10^{11}-10^{11.9}\, L_{\odot}$ and  $L_{IR}\geq 10^{12}\, L_{\odot}$, respectively. One prototype of merging galaxies is the object NGC\,6240, being one of the most studied merger system in all wavelengths ranges \citep[e.g.][]{fosbury79,wright84,joseph84,max05,max07,engel10,meijerink13,feruglio13,mori14,hagiwara15,tunnard15}. NGC\,6240 is located at a distance of 97\,Mpc, has two nuclei separated by $\sim$1\farcs5 (corresponding to $\sim700$\,pc) and each of them hosts an AGN as detected in X-ray wavelengths \citep{komossa03}. 

 NGC\,6240 presents an infrared luminosity of L$_{IR}=10^{11.8}\,$L$_{\odot}$ and lies at the limit between LIRGs and ULIRGs \citep{sanders96}. It presents the strongest H$_2$ line emission in the near-IR observd for this kind of objects, with a luminosity $L_{\rm H_2}\sim 10^9 L_{\odot}$ \citep[e.g.][]{joseph84}. Then, NGC\,6240 is an ideal object to study the origin of the molecular hydrogen emission in merger systems with AGNs, as the H$_2$ line emission can be originated both by Starburts and AGNs. Several studies mapped the H$_2$ emission in the inner region of NGC\,6240 \citep[e.g.][]{tanaka91,werf93,sugai97,tecza00,ohyama03,bogdanovic03,max05,engel10}, however its origin is still not well understood.  Besides the strong H$_2$ line emission, NGC\,6240 shows also emission from the ionized gas \citep[e.g.][]{engel10}. The understanding of the excitation and ionization conditions is a fundamental key to better understand the formation/evolution history of this object and galaxy evolution.

In this work we use public near-infrared Integral Field Spectroscopic data from the Gemini Observatory Archive (projects GN-2007A-Q-62 and GN-2008A-Q-33 -- PI: M. Tecza) to map the flux distributions and kinematics for the molecular and ionized gas and investigate the excitation mechanisms of the H$_2$ line emission. These data were originally observed with the aim of measure the mass of the super-massive black holes at the centers of both nuclei of NGC\,6240 and comprises observations at the K-band obtained with the Near-Infrared Integral Field Spectrograph (NIFS) at the Gemini North telescope. In this work we map and discuss the emission-line flux distributions and kinematics for the molecular and ionized gas, traced by H$_2$, H\,{\sc i}, [Fe\,{\sc ii}] and [Ca\,{\sc viii}] emission lines.  

This paper is organized as follows. Section 2 presents a description of the observations and data reduction, the emission line flux and kinematic maps are presented in section 3 and are discussed in section 4. Finally, a summary of this work is presented in Section 5.

\section{Observations and Data Reduction}

 The integral field spectroscopic data of NGC\,6240 were obtained from Gemini Observatory Archive\footnote{http://archive.gemini.edu}. The data were obtained in the near-IR K-band using the Gemini NIFS \citep{mcgregor03} operating with the ALTAIR adaptive optics module. The observations were done in June, 2007 and April, 2008 under the projects IDs GN-2007A-Q-62 and GN-2008A-Q-33 (PI: M. Tecza), respectively. The observations comprised one single exposure of 300\,s centred at the southern nucleus of NGC\,6240 and 16 exposures of 300\,s centred at its northern nucleus.

 The data reduction procedure was done using the {\sc gemini.nifs} IRAF package and followed standard spectroscopic data reduction procedures \citep[e.g.][]{n4051,mrk1157,schonell14}, including flat fielding, sky subtraction, wavelength calibration, spatial distortion correction and telluric absorption cancellation. The spectra have been flux calibrated by interpolating a black-body function to the telluric standard star and the final data cube has been constructed by average combining the data cubes from individual exposures, taking into account the offset positions. 
The final data cube covers the inner 3\farcs0$\times$5\farcs15 (1.41$\times$2.35 kpc$^{2}$) of NGC\,6240 and contains about 6000 spectra. The angular resolution of the final cube is 0\farcs19, as derived from the full width at half maximum of the continuum emission of a telluric standard star, corresponding to 90\,pc at the galaxy.

\section{Results}
\label{results}

As can be seen in Fig.\ref{spec}, the K-band spectra of NGC\,6240  present several emission lines, most of them of the molecular hydrogen. The following emission lines were detected in the NIFS spectra: H$_2$ lines at 2.0338, 2.1218, 2.1542, 2.2014, 2.2235 and 2.2477\,$\mu$m, Br$\gamma$, [Fe\,{\sc ii}]$\lambda2.0719\,\mu$m, He\,{\sc i}$\lambda$\,2.0587\,$\mu$m and [Ca\,{\sc viii}]$\lambda$2.3210\,$\mu$m.  The CO absorption band heads at 2.3\,$\mu$m are also clearly present. The spectra shown in Fig.~\ref{spec} were obtained for a circular aperture of 0\farcs25 diameter centred at the locations indicated in Fig.~\ref{fluxmaps}. 

\subsection{Emission-line flux distributions}

\begin{figure}[!t]
  \includegraphics[width=\columnwidth]{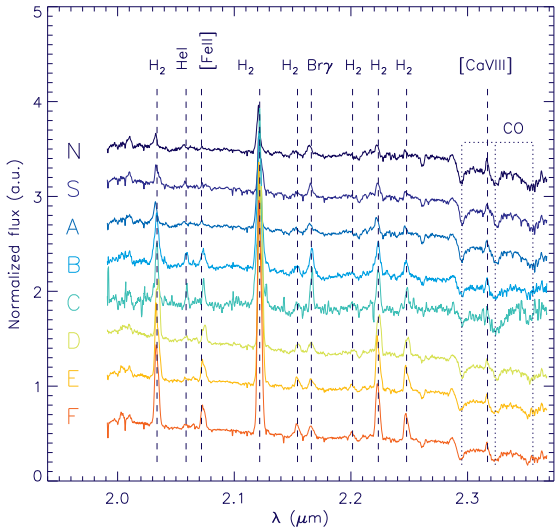}
  \caption{Sample spectra obtained for a circular aperture with 0\farcs25 diameter for the positions marked in Fig.~\ref{fluxmaps}, with the emission/absorption lines identified.}
  \label{spec}
\end{figure}

\begin{figure*}[!t]
  \includegraphics[scale=0.68]{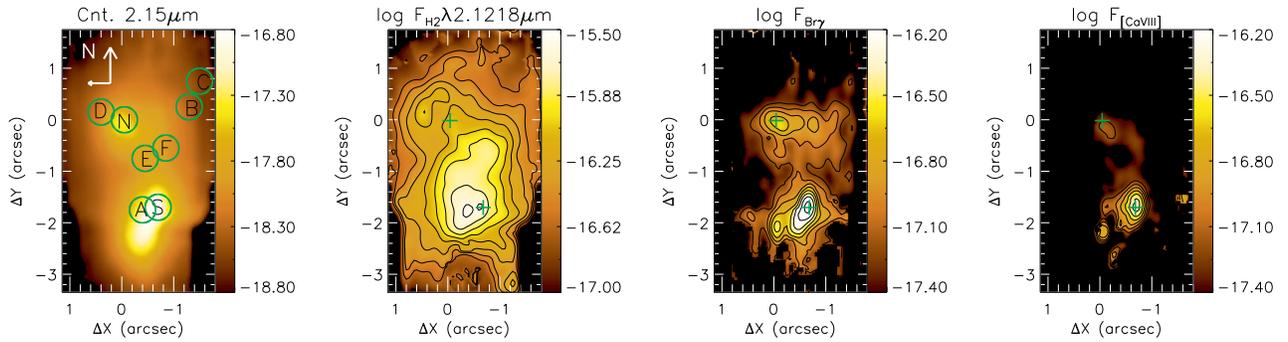}
  \caption{Continuum map and flux distributions for the $H_2\,\lambda2.12\,\mu$m, Br$\gamma$ and [Ca\,{\sc viii}]$\lambda$2.32\,$\mu$m emission lines. Black regions represent masked locations, where the S/N was not high enough to fit the line profiles. The circles mark the locations where the spectra of Fig.~\ref{spec} were extracted and the color bars show the fluxes in logarithmic units [log erg\,s$^{-1}$\,cm$^{-2}$].}
  \label{fluxmaps}
\end{figure*}

In order to map the emission-line flux distribution and gas kinematics, we used the {\sc profit} routine \citep{profit} to fit the  profiles of the  H$_2\lambda$2.1218\,$\mu$m and  Br$\gamma$ emission lines by Gauss-Hermite series. The H$_2$2.1218$\,\mu$m was chosen because it presents the highest signal-to-noise ratio among the H$_2$ lines and all lines show similar flux distributions and kinematics. As already known, the near-IR emission lines present complex profiles \citep[e.g.][]{engel10} and the use of Gauss-Hermite series allow us to map deviation from a simple Gaussian curve and better reproduce the line profiles, thus get better estimates for their fluxes and kinematics.  The resulting flux maps are shown in Fig.~\ref{fluxmaps}. We do not present maps for the  [Fe\,{\sc ii}] and He\,{\sc i}, because they are detected only at some locations. The  [Ca\,{\sc viii}]$\lambda$2.3210\,$\mu$m flux map was obtained directly by integration of the emission-line profile, after the subtraction of the contribution of the underlying stellar population, by the fitting of the CO absorption band-heads using the pPXF code \citep{ppxf} and selected stellar spectra from \citet{winge09}. In this figure, black regions represent masked locations, where the signal-to-noise ratio was not high enough to fit the line profiles.

The left panel of Fig.~\ref{fluxmaps} show the K-band continuum obtained by calculating the average of the fluxes within a spectral window of 100\,\AA, centred at 2.15\,$\mu$m (a region free of emission lines).  The location of the northern and southern nuclei were defined as the peak of the continuum emission for each nuclei, and marked by + sign in all maps. The green circles mark the locations and apertures used to extract the spectra shown in Fig.~\ref{spec}. The H$_2$ shows emission in the whole field of view (FoV) and presents its flux peak at 0\farcs4 southeast of the southern nucleus and enhanced emission in a region between both nuclei. Additionally it is observed an  `arc' structure surrounding the northern nucleus. The Br$\gamma$ flux distribution is well correlated with the continuum emission, presenting the two peaks at the locations of the nuclei and lower emission at locations between them. Another interesting structure observed in the Br$\gamma$ map is an extended emission seen to the west of the northern nucleus. Finally, the coronal line of [Ca\,{\sc viii}] presents emission at both nuclei and at the region between them. The emission peak is observed at the southern nucleus, similarly to the Br$\gamma$ emission.

In Tables \ref{tab:flux1g} and \ref{tab:flux2g} we present the fluxes of the emission lines obtained by fitting them at the locations identified at the continuum map of Fig.~\ref{fluxmaps}. The spectra were extracted within circular apertures of 0\farcs25.  These particular locations were selected to represent typical spectra of the galaxy at distinct locations: positions ``S" and ``N", correspond to the locations of the southern and northern nuclei, position ``A" is centred at the location of the peak of H$_2$ emission, positions ``B" and ``C" represent locations where  the Br$\gamma$ flux map shows an extended structure to the west of the northern nucleus, near the border of the FoV. Position ``D" corresponds to a location where the H$_2$ present enhanced emission next to the northern nucleus, while positions ``E" and ``F" are locations between both nuclei. As already mentioned above, the emission-line profiles at some locations are complex and cannot be reproduced by a single Gaussian. For the locations ``N", ``D", ``E" and ``F", we fitted the emission line-profiles by two gaussians, while for the other locations a single Gaussian curve produces a good representation of the line profiles.  Table \ref{tab:flux1g} shows the fluxes obtained for the locations where a single Gaussian was fitted to the emission-line profiles and table~\ref{tab:flux2g} shows the fluxes for locations which two gaussians were needed to reproduce each line profile. The fluxes are shown for the blue and red component. Average velocities are shown in Table~\ref{tab:pars} and will be discussed in Section~\ref{disc}.

\begin{table*}[!t]\centering
   \caption{Emission-line fluxes obtained by fitting the line profiles by a single Gaussian for the spectra shown in Fig.\ref{spec}. The fluxes are shown in units of 10$^{-16}$ erg\,s$^{-1}$\,cm$^{-2}$.} \label{tab:flux1g}
 \begin{tabular}{lcccc}
\hline
           Line             & S 	       & A		  & B	    & C        \\
\hline
      $H_2\lambda$2.0338    &  32.1$\pm$5.4    &  43.9$\pm$4.2    &  9.4$\pm$0.4    &	  5.1$\pm$0.7 \\
  He{\sc i}$\lambda$2.0587  &  8.7$\pm$5.0     & 5.8$\pm$3.7	  & 1.6$\pm$0.3     &  0.9$\pm$0.3    \\
 $[Fe II]\lambda$2.0719     &	4.3$\pm$3.3    &  6.2$\pm$2.6	  & 2.8$\pm$0.4     & 1.3$\pm$0.4     \\
     $H_2\lambda$2.1218     &	147.5$\pm$6.5  &   175.9$\pm$5.0  &   27.7$\pm$0.4  &  19.6$\pm$0.8   \\
      $H_2\lambda$2.1542    &	19.1$\pm$10.4  & 18.1$\pm$7.4	  & 1.6$\pm$0.4     &	--	      \\	  
    HI$\lambda$2.1662       &  45.5$\pm$9.1    &  34.7$\pm$7.7    &  5.2$\pm$0.4    & 3.5$\pm$0.7    \\
     $H_2\lambda$2.2014     & 22.4$\pm$13.7    &   --		  &	--	    &	--	     \\
      $H_2\lambda$2.2235    & 32.5$\pm$6.6     &  50.5$\pm$6.5    & 6.4$\pm$0.4     & 3.4$\pm$0.7    \\
      $H_2\lambda$2.2477    & 16.5$\pm$4.7     &  15.7$\pm$3.6    &  2.9$\pm$0.4    &	1.4$\pm$0.6  \\
 $[CaVIII]\lambda$2.3211    &  32.8$\pm$4.3    &  22.5$\pm$3.7    &  1.1$\pm$0.2    &	--	     \\
\hline
  \end{tabular}
\end{table*}

\begin{table*}[!t]\centering
  \caption{Same as Table~\ref{tab:flux1g}, but for locations where two gaussians are needed to reproduce the line profiles. The fluxes are shown in units of 10$^{-16}$ erg\,s$^{-1}$\,cm$^{-2}$. The fluxes for the [Ca\,{\sc viii}] was obtained by fitting a single gaussian, due to its low-signal-to-noise ratio.} \label{tab:flux2g}
\label{tab:flux2g}
 \begin{tabular}{lrcccc}
    \hline
         Line               & Comp. & N 	   & D  	    & E 	     & F	  \\
    \hline
      $H_2\lambda$2.0338    & blue  & 11.8$\pm$1.1 &   8.0$\pm$0.2  &	 21.7$\pm$0.4	 &  30.0$\pm$0.5 \\
                            & red   &  8.0$\pm$1.5 &   12.3$\pm$0.3 &	 15.9$\pm$.5	 &   7.5$\pm$0.3 \\
  He{\sc i}$\lambda$2.0587  & blue  &	   -	   &   0.5$\pm$0.4  &	 0.5$\pm$0.2	 &	-	 \\
                            & red   &	   -	   &   1.2$\pm$0.4  &	 0.9$\pm$0.3	 &	-	 \\
 $[Fe II]\lambda$2.0670     & blue  &	   -	   &   0.7$\pm$0.3  &		-	 &	 -	 \\ 
                            & red   &	    -	   &   0.5$\pm$0.2  &		-	 &	-	 \\ 
 $[Fe II]\lambda$2.0719     & blue  &	   -	   &   1.9$\pm$0.2  &	    6.1$\pm$0.4  & 6.8$\pm$0.4   \\
                            & red   &	   -	   &   4.3$\pm$0.3  &	    5.0$\pm$0.5  & 4.6$\pm$0.4   \\
     $H_2\lambda$2.1218     & blue  & 29.5$\pm$1.0 &   21.5$\pm$0.2 &	   56.7$\pm$0.4  & 74.2$\pm$0.4  \\
                            & red   & 28.9$\pm$1.6 &   38.7$\pm$0.3 &	   56.0$\pm$0.6  & 36.5$\pm$0.4  \\
      $H_2\lambda$2.1542    & blue  &	  -	   &   1.9$\pm$0.3  &	   3.7$\pm$0.5   & 2.9$\pm$0.4   \\
                            & red   &	  -	   &   1.9$\pm$0.3  &	   2.4$\pm$0.6   & 1.7$\pm$0.4   \\
    HI$\lambda$2.1662       & blue  & 12.1$\pm$1.6 &   1.9$\pm$0.4  &	   4.6$\pm$0.6   &  3.0$\pm$0.5  \\  
                            & red   &  1.5$\pm$0.5 &   2.5$\pm$0.4  &	   2.1$\pm$0.6   &  2.6$\pm$0.6  \\  
     $H_2\lambda$2.2014     & blue  &	  -	   &	     -      &	2.2$\pm$0.5	 &  2.0$\pm$0.4 \\
                            & red   &	  -	   &	     -      &	0.2$\pm$0.1	 &  0.6$\pm$0.3 \\
      $H_2\lambda$2.2235    & blue  & 9.9$\pm$1.3  &   6.6$\pm$0.3  &	  16.1$\pm$0.4   &18.3$\pm$0.5  \\
                            & red   & 4.5$\pm$1.4  &   8.3$\pm$0.3  &	  10.8$\pm$0.5   & 9.4$\pm$0.5  \\
      $H_2\lambda$2.2477    & blue  & 5.5$\pm$1.1  &   3.3$\pm$0.3  &	   7.2$\pm$0.4   & 9.2$\pm$0.5  \\
                            & red   & 5.1$\pm$1.8  &   4.1$\pm$0.3  &	   6.4$\pm$0.6   & 3.8$\pm$0.4  \\
 $[CaVIII]\lambda$2.3211    & total &  7.3$\pm$1.3 &   1.5$\pm$0.2  &	   2.3$\pm$0.3   &  1.9$\pm$0.3 \\
    \hline
  \end{tabular}
\end{table*}

\subsection{Gas and stellar kinematics}

\begin{figure*}[!t]
  \includegraphics[scale=0.82]{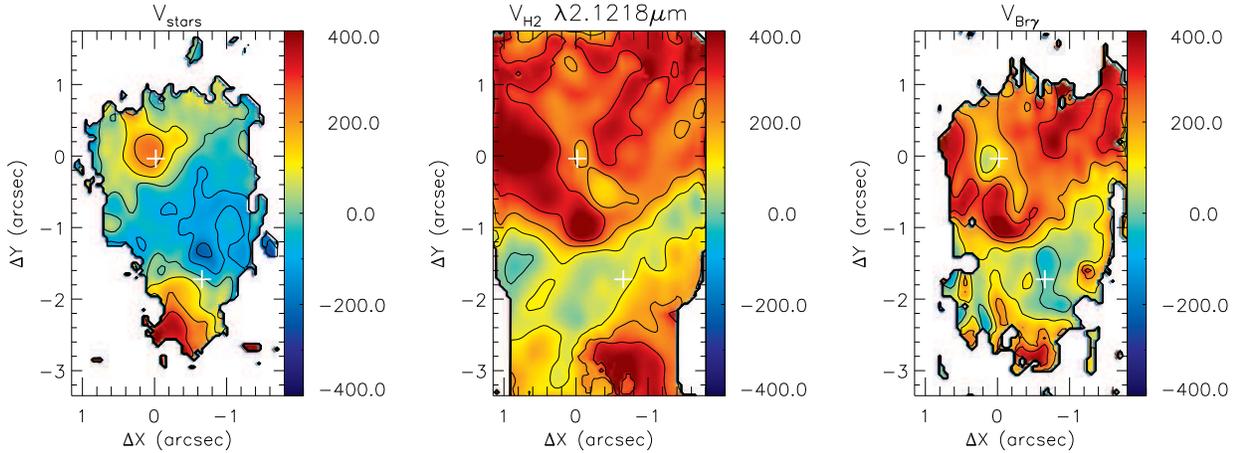}
  \caption{Stellar and gas velocity fields. White regions represent masked locations, where the S/N was not high enough obtain good fits. The color bars show show the velocities in units of km\,s$^{-1}$, relative to the systemic velocity of the southern nucleus.}
  \label{velmaps}
\end{figure*}

\begin{figure*}[!t]
  \includegraphics[scale=0.82]{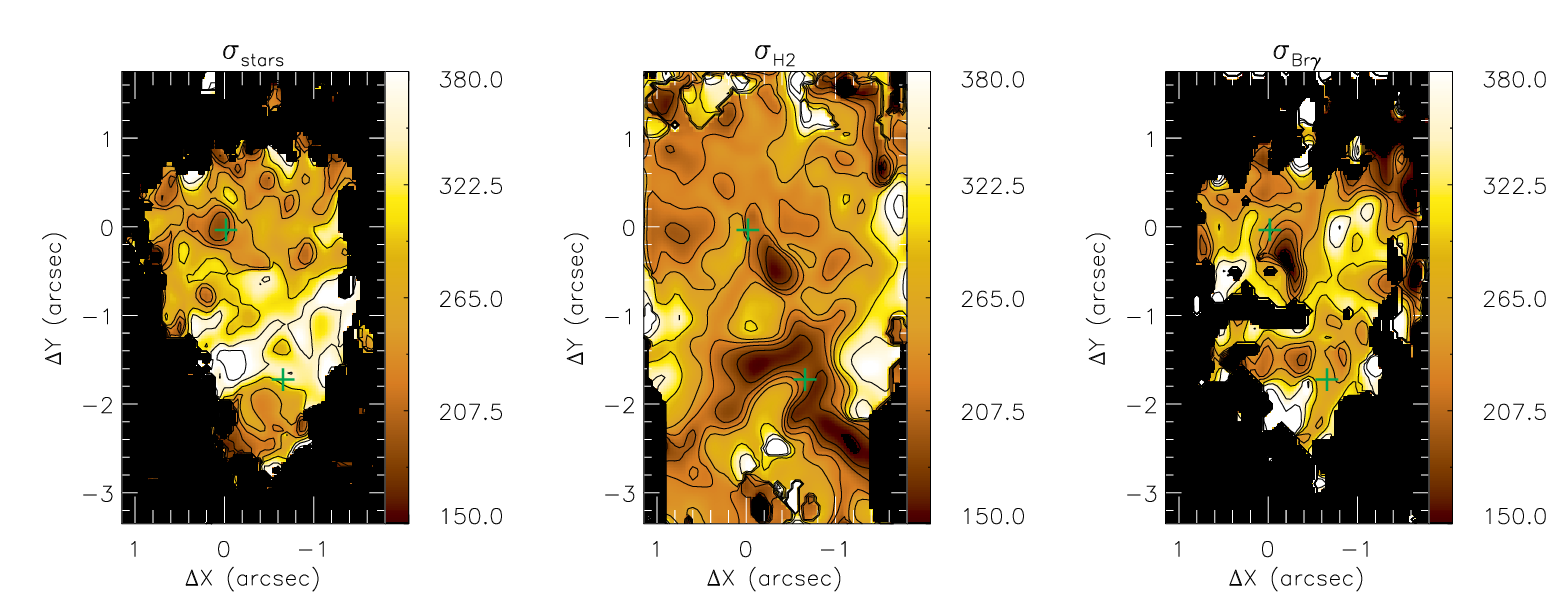}
  \caption{Stellar and gas velocity dispersion maps. The color scale is shown in units of km\,s$^{-1}$.}
  \label{sigmaps}
\end{figure*}

In Figure~\ref{velmaps} we present the velocity fields for the stars, H$_2$ and Br$\gamma$ emitting gas. The stellar velocity field was obtained by the fitting of the CO absorptions band heads at $\sim$2.3\,$\mu$m using the Penalized Pixel-Fitting (pPXF) routine \citep{ppxf}, that fits the stellar kinematics assuming that the line-of-sight velocity distribution (LOSVD) of the stars is well reproduced by Gauss-Hermite series. The best fit of the galaxy spectrum is obtained by convolving a template spectra with a given LOSVD and the pPXF outputs the stellar centroid velocity ($V_*$), velocity dispersion ($\sigma_*$) and the higher-order Gauss-Hermite moments $h_{3*}$ and $h_{4*}$. In this work, we 
have used template spectra from the  Gemini library of late spectral type stars observed with the Gemini Near-Infrared Spectrograph (GNIRS) Integal Field Unit (IFU) and  
NIFS \citep{winge09}, as these spectra have similar spectral resolution of the NGC\,6240 data. 

The stellar velocity field is shown at the left panel of Fig.~\ref{velmaps}, from which we subtracted the systemic velocity  ($Vs=7167$\,km\,s$^{-1}$) of the southern nucleus, as obtained by the fitting of nuclear spectrum integrated within an circular aperture with radius of 0\farcs25, using the pPXF code. The stellar velocity field clearly shows the rotation pattern for two disks, with kinematical centers at the locations of both nuclei. The velocity amplitude for the southern nucleus is about 400~\,km\,s$^{-1}$ with the major axis of the disk oriented along the position angle $PA\approx-30^\circ$. The velocity amplitude for the northern disk is about 200~\,km\,s$^{-1}$ and its major axis is oriented along $PA\approx45^\circ$. 

The molecular and ionized gas velocity fields (middle and left panels of Fig.~\ref{velmaps}) are very distinct than that of the stars. The velocities range from $-200\,$km\,s$^{-1}$ at locations next to the southern nucleus to  $400\,$km\,s$^{-1}$ to the east and northwest of the northern nucleus.  The gas velocity fields do not present clear evidence of an  organized disk rotation, showing complex kinematics.

Figure~\ref{sigmaps} shows the stellar, $H_2$ and Br$\gamma$ velocity dispersion ($\sigma$) maps. All maps show high $\sigma$ values of up to 350\,km\,s$^{-1}$, but with distinct distributions. The highest $\sigma$ values for the stars are observed at locations between both nuclei and the lowest value of  150\,km\,s$^{-1}$  is observed for the northern nucleus. The H$_2$ and Br$\gamma$ maps show complex distributions with high and low $\sigma$ values observed at several locations, revealing the complexity of the emission line profiles and gas kinematics of NGC\,6240.

\section{Discussion}\label{disc}

\subsection{Kinematics}

The stellar and molecular gas kinematics of the inner region of NGC\,6240 have already been investigated using near-IR IFS \citep[e.g.][]{engel10,medling11}. \citet{engel10} present spatially resolved K-band IFS obtained with the SINFONI instrument at the Very Large Telescope (VLT) at a spatial resolution of 60~pc and $^{12}$CO$(J=2-1)$ line observations using the IRAM millimeter interferometer at a spatial resolution of 175~pc for the inner 2\farcs45$\times$3\farcs95 of NGC\,6240. They show that the H$_2$ and CO lines present similar flux distributions and kinematics, showing similar complex line profiles and are distinct than those of the stars. The stellar and molecular gas kinematics and flux distributions presented here are very similar to that shown in \citet{engel10}, however the FoV of the NIFS data is larger (3\farcs0$\times$5\farcs15) and we were able to fit the H$_2$ line profile over the whole region. In particular, the NIFS H$_2$ velocity field reveals a region of redshifts at 1$^{\prime\prime}$ south of the southern nucleus. 

 \citet{engel10} used the kinemetry method \citep{kinemetry} to parametrize the stellar velocity field of NGC\,6240 and determine the position of the kinematical centers of both nuclei to compare with the location of the AGNs, as determined by near-IR, radio and X-rays images \citep{max07}. They found that  position of the kinematical center of the northern nucleus and the location of the AGN are coincident, while for the southern nucleus a separation of 0\farcs22 is observed between the locations of the kinematical center and of the AGN. \citet{medling11} also used K-band adaptive optics IFS to map the stellar kinematics from the CO absorption band heads with the observations performed with the  OH-Suppressing InfraRed Imaging Spectrograph (OSIRIS) on the Keck II telescope. These authors combined the kinematics with high resolution images obtained with the Near InfraRed Camera 2 (NIRC2) at the same telescope to measure the mass of the super-massive black hole (SMBH) of the southern nucleus of NGC\,6240. By modeling the stellar kinematics using a Jeans Axisymmetric Multi-Gaussian mass model to reproduce the observed velocity dispersion and a Keplerian rotating disk to model the velocity field, they found that the mass of the SMBH is in the range $(0.87-2.0)\times10^{9}$~M$_\odot$, being consistent with the value obtained from the $M_{\bullet}-\sigma$ relation. The stellar kinematics derived from the NIFS data is consistent with that obtained by \citet{medling11} and \citet{engel10}.

As already mentioned above, the H$_2$ velocity field and $\sigma$ map are similar to that presented by \citet{engel10} and suggest that the H$_2$ emission arises from a turbulent gas, instead from rotating disks, as commonly observed for isolated active galaxies \citep[e.g.][]{mrk79,n4051}.  Both velocity maps show complex kinematics and some diferences are observed between the velocity fields of H$_2$ and Br$\gamma$, in particular for the southern nucleus. Distinct kinematics and flux distributions is commonly observed for active galaxies using similar IFS data \citep[e.g.][]{diniz15,mrk79,mrk1066}. 


\subsection{The origin of the H$_2$ emission}

The $H_2$ line emission in AGNs can be originated by (i) thermal processes, as heating of the gas by shocks \citep{hollembach89} or by X-rays from the central AGN  \citep{maloney96} or (ii) by excitation of the gas due to infrared fluorescence through absorption of UV photons \citep{black87}. 
NGC\,6240 shows the strongest near-IR H$_2$ emission already observed with a luminosity $L_{\rm H_2}\sim 10^9 L_{\odot}$ \citep[e.g.][]{joseph84} and several studies have been addressed to study the H$_2$ emission. \citet{engel10} present high resolution (0\farcs13) IFS data of NGC\,6240 and compare with interferometric CO(2-1) line observations. They found that the H$_2$ and CO show similar line profiles and conclude that the molecular gas emission arises from a highly disturbed gas, possible due to shocks. Other studies spectroscopic and imaging also support that the main excitation mechanism of the near-IR H$_2$ lines are shocks \citep[e.g.][]{werf93,sugai97,tecza00,ohyama03,max05}. However, fluorescent excitation of the H$_2$ lines through absorption of soft-UV photons (912--1108 \AA) in the Lyman and Werner bands may also be present \citep[e.g.][]{tanaka91,bogdanovic03}.

In order to distinguish between thermal and fluorescent  excitation we can use the rotational and vibrational temperatures of the H$_2$ emitting gas. For fluorescent excitation the vibrational temperature is expected to be high ($\gtrsim6000$\,K 
-- non-local UV photons overpopulate the highest energy levels) and the rotational temperature should be low. On the other hand, for thermal excitation both temperatures are similar.
The vibrational temperature can be found by 
\[T_{\rm vib}\cong {\rm 5600/ ln\left(1.355\,\frac{F_{H_{2}\lambda2.1213}}{F_{H_{2}\lambda2.2471}}\right)}\]  and 
the rotational temperature by 
\[T_{\rm rot} \cong -{\rm 1113/ln\left(0.323\,\frac{F_{H_{2}\lambda2.0332}}{F_{H_{2}\lambda2.2227}}\right)},\] 
where $F_{H_{2}\lambda i}$ are the fluxes of H$_2\lambda i$ lines  \citep{reunanen02}. Using the fluxes from Tables~\ref{tab:flux1g} and \ref{tab:flux2g} together with the equations above, we estimated $T_{\rm vib}$ and $T_{\rm rot}$ for the locations identified at Fig.~\ref{fluxmaps}. The corresponding temperatures are shown in Table~\ref{tab:pars}. For regions where the profiles were fitted by two Gaussians we show the temperatures for each  component, as well as their centroid velocity (relative to the systemic velocity)  and the $\sigma$. At each location, we constrained the distinct H$_2$ lines to have the same kinematics during the fitting of the profiles. At all locations both temperatures are small ($\sim1000-2500\,$K), being consistent with H$_2$ excitation by thermal processes.

The H$_2\,\lambda2.12\,\mu$m/Br$\gamma$ emission-line ratio can be used to further study the excitation mechanism of the H$_2$ emission lines. Using long-slit spectra of the nuclear region of AGNs and non-active galaxies it has been concluded that typical values for this ratio are: H$_2\,\lambda2.12\,\mu$m/Br$\gamma < 0.6$ for Starburts, $0.6<$H$_2\,\lambda2.12\,\mu$m/Br$\gamma<2.0$ for Seyfert galaxies and higher values are expected when shocks are important \citep[e.g.][]{ardila05,ardila04,reunanen02}. However, more recently, \citet{rogerio13} used a sample of  67 AGN and Star-forming galaxies (SFG) using spectra obtained with the Infrared Telescope Facility SpeX at the near-IR, together with  photo-ionization models to investigate the origin of the H$_2$ emission. Their sample included not only Seyfert galaxies, but also Low-Ionization Nuclear Emission Regions (LINERs) and they found that typical values for AGN are in the range $0.4<$H$_2\,\lambda2.12\,\mu$m/Br$\gamma<6.0$, with the highest values observed for LINERs.

IFS has also been used to investigate origin of the H$_2$ emission  in AGNs and LIRGS \citep[e.g.][]{colina15,n5929,n1068-exc,mrk1066-exc,dors12}. A very detailed study is presented by \citet{colina15} using near-IR IFS with the SINFONI instrument at the Very Large Telescope (VLT) to discuss the[Fe\,{\sc ii}]\,1.64$\mu$m/Br$\gamma$ vs. H$_2\,\lambda2.12\,\mu$m/Br$\gamma$ diagnostic diagram.  They found the following limits for log(\feii$\lambda$1.64/\br) and log(\hm/\br): AGN-dominated:$[-0.3,~+1.5]$ and $[-0.3,~+0.9]$; young-stars dominated: $[-0.4,~+0.4]$ and $[-1.2,~-0.1]$ and Supernovae(SNe)-dominated: $[+0.2,~+1.2]$ and $[-0.4,~+0.4]$.

\begin{figure*}[!t]\centering
  \includegraphics[scale=0.95]{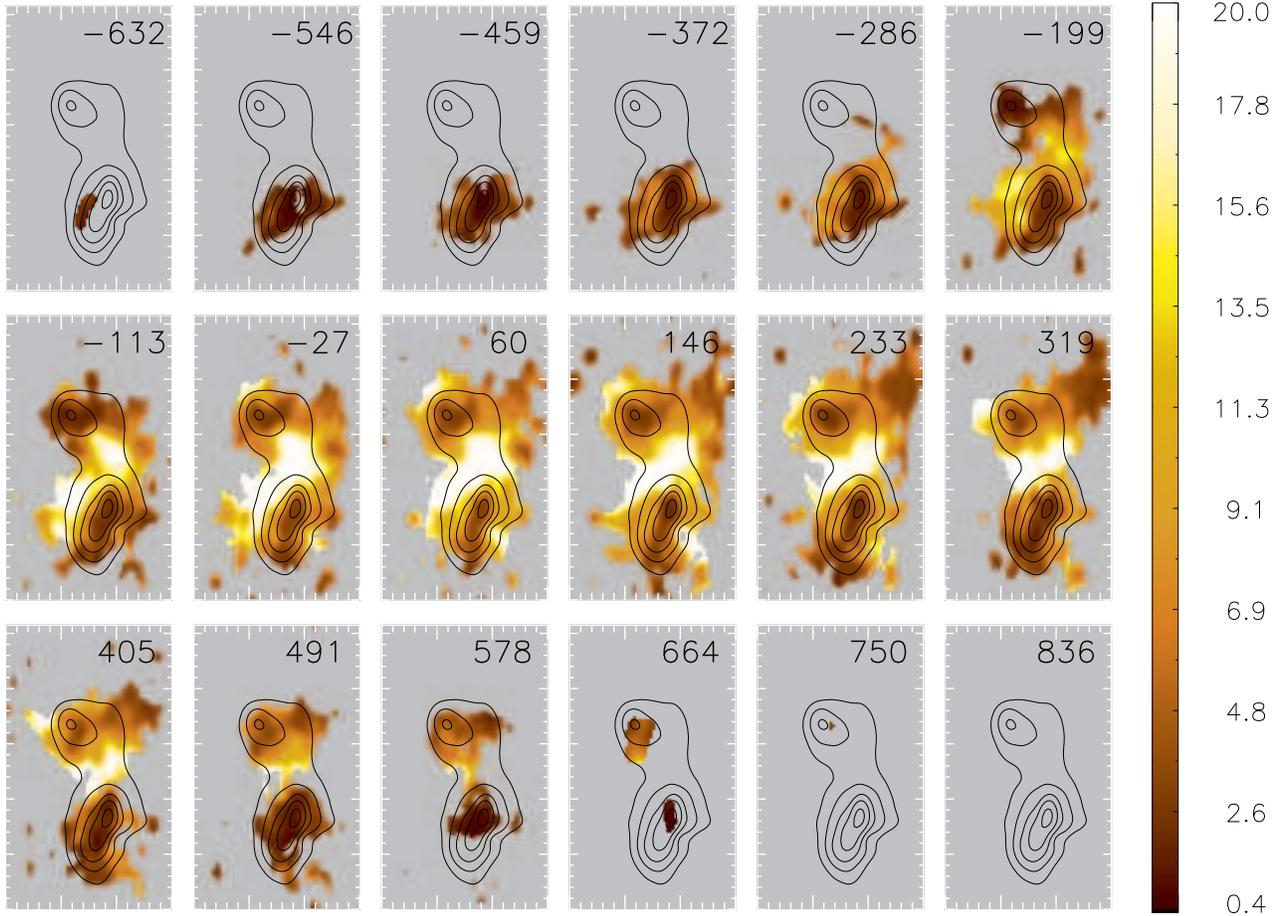}
  \caption{H$_2\,\lambda2.12\,\mu$m/Br$\gamma$ channel maps for a velocity bin of 90~km\,s$^{-1}$, corresponding to three spectral pixels. The corresponding velocity is shown in the top-right corner of each panel in units of km\,s$^{-1}$, relative to the velocity of the stellar velocity of the northern nucleus.}
  \label{slices}
\end{figure*}

As already mentioned above, the H$_2$ emission-line profiles are complex at many locations of the inner region of NGC\,6240. To investigate the H$_2$ line excitation at distinct locations and velocities, we constructed  H$_2\,\lambda2.12\,\mu$m/Br$\gamma$ line-ratio channel maps by mapping this line ratio at velocity bins. The resulting maps were integrated within velocity bins of bin of 90~km\,s$^{-1}$, (corresponding to three spectral pixels) and are shown in Fig.~\ref{slices}. The central velocity of each bin relative to the systemic velocity of the southern nucleus is shown at the top-right corner of each panel in units of km\,s$^{-1}$. Regions with fluxes in one or both emission lines smaller than 3 times the standard deviation of the continuum next to the line were masked out in order to avoid spurious features. These regions are shown in gray, while the black contours are from the H$_2$ flux map of Fig.~\ref{fluxmaps}. 

As seen in Fig.~\ref{slices} the H$_2\,\lambda2.12\,\mu$m/Br$\gamma$ ranges from values close to zero to values of up to 20. At velocities from $\sim-600$ to $-200$\,km\,s$^{-1}$, most of the line emission arises from the southern nucleus and $0.4<$H$_2\,\lambda2.12\,\mu$m/Br$\gamma<2.0$ ratio. At the highest redshifts ($> 400$\,km\,s$^{-1}$), similar ratios are observed for the southern nucleus and at lower velocities the southern nucleus shows values smaller than 6.  The northern nucleus is seen in velocity bins from $-200$ to $600$\,km\,s$^{-1}$ and show H$_2\,\lambda2.12\,\mu$m/Br$\gamma$  values between 2.0 and 6.0. As discussed in \citet{rogerio13} and \citet{dors12} values of $0.4<$H$_2\,\lambda2.12\,\mu$m/Br$\gamma<6.0$ are typical values for AGNs and can be explained by emission of gas heated by X-rays emitted from the central engine. Thus, we conclude that H$_2$ emission from both nuclei is dominated by emission of gas excited by X-rays. 

The highest  H$_2\,\lambda2.12\,\mu$m/Br$\gamma$ values are seen at lower velocities ($-200$ to $400$\,km\,s$^{-1}$) and observed in the region between the nuclei. Such high values suggest that shocks are 
 playing an important role in the excitation of the H$_2$ lines from this region, as discussed above. Finally, at velocities from 230 to 320\,km\,s$^{-1}$ a region with H$_2\,\lambda2.12\,\mu$m/Br$\gamma\sim4$ is observed at 1\farcs2 northwest from the northern nucleus.  As this is an extra-nuclear region and the values are smaller than that seen at locations between the nuclei, the H$_2$ emission may have an additional component due to fluorescent excitation.

In Table~\ref{tab:pars} we show the H$_2\,\lambda2.12\,\mu$m/Br$\gamma$ values obtained for the locations identified in Fig.~\ref{fluxmaps}.  It can be seen that regions where a single Gaussian component represent a  good model of the line profiles, the above line ratio is smaller than regions where the profiles are more complex (modeled by two Gaussians). This result supports the above conclusion that shocks play an important role in locations where the gas present a more disturbed kinematics. At the northern nucleus, the line profile show two components, with the blue component being due to X-ray excitation and the red component due to shocks, as suggested by the line ratios.

\begin{table*}[!t]\centering
  \caption{Vibrational and rotational temperatures (cols 2 and 3), $\frac{H_2\lambda2.122}{Br\gamma}$ and $\frac{[FeII]\lambda2.071}{Br\gamma}$ line ratios (cols 4 and 5) and centroid velocity relative to the systemic velocity and velocity dispersions (cols 6 and 7) for the locations identified in Fig.~\ref{fluxmaps} (col 1). } 
\label{tab:pars}
 \begin{tabular}{ccccccc}
\hline
         Pos   & $T_{\rm vib}$ (K) & $T_{\rm rot}$ (K)& $\frac{H_2\lambda2.122}{Br\gamma}$ & $\frac{[FeII]\lambda2.071}{Br\gamma}$& $V$ (km\,s$^{-1}$)& $\sigma$ (km\,s$^{-1}$)  \\
\hline
     S         &  2245$\pm$203  &   974$\pm$ 37  &  3.2$\pm$0.7  &  0.1$\pm$0.1  & $-58\pm20$	     & $470\pm48$     \\   
     A         &  2058$\pm$144  &   876$\pm$ 26  &  5.1$\pm$1.1  &  0.2$\pm$0.1  & $-54\pm12$	     & $543\pm56$      \\   
     B         &  2187$\pm$102  &  1492$\pm$ 43  &  5.3$\pm$0.4  &  0.5$\pm$0.1  & $139\pm17$	     & $591\pm30$     \\
     C         &  1902$\pm$229  &  1535$\pm$198  &  5.6$\pm$1.1  &  0.4$\pm$0.1  & $187\pm16$	     & $383\pm38$     \\						      
     N (blue)  &  2823$\pm$229  &  1166$\pm$ 54  &  2.4$\pm$0.3  &  -            & $31\pm7$          & $346\pm22$   \\ 
     N (red)   &  2747$\pm$381  &  2006$\pm$454  & 19.3$\pm$6.5  &  -            & $410\pm27$        & $407\pm32$   \\
     D (blue)  &  2571$\pm$ 95  &  1186$\pm$ 27  & 11.3$\pm$2.4  &  1.0$\pm$0.2  & $40\pm15$         & $361\pm24$  \\ 
     D (red)   &  2197$\pm$ 55  &  1510$\pm$ 25  & 15.5$\pm$2.5  &  1.7$\pm$0.3  & $352\pm18$        & $397\pm27$   \\						      
     E (blue)  &  2365$\pm$ 47  &  1338$\pm$ 10  & 12.3$\pm$1.6  &  1.3$\pm$0.2  & $35\pm6$          & $341\pm17$  \\ 
     E (red)   &  2264$\pm$ 74  &  1497$\pm$ 31  & 26.7$\pm$7.6  &  2.4$\pm$0.7  & $381\pm18$        & $380\pm40$   \\						      
     F (blue)  &  2341$\pm$ 47  &  1750$\pm$ 30  & 24.7$\pm$4.1  &  2.3$\pm$0.4  & $29\pm10$         & $387\pm13$  \\ 
     F (red)   &  2182$\pm$ 78  &   820$\pm$  8  & 14.0$\pm$3.2  &  1.8$\pm$0.4  & $346\pm13$        & $358\pm22$   \\						      
\hline
  \end{tabular}
\end{table*}

\subsection{The origin of the [Fe\,{\sc ii}] emission}

Although the K-band spectra of NGC\,6240 show only weak [Fe\,{\sc ii}] line emission, we can speculate on its origin. The ratio between the  [Fe\,{\sc ii}] and H recombination lines (as for example \feii$\lambda$1.64/\br)  can be used to distinguish between shocks and X-ray excitation of the [Fe\,{\sc ii}] lines \citep[e.g.][]{colina15}.  In Table~\ref{tab:pars} we show the {[FeII]$\lambda2.071$/Br$\gamma$
 ratio values for the locations identified in Fig.~\ref{fluxmaps}. It can be seen that this ratio follows the same trend of the H$_2\,\lambda2.12\,\mu$m/Br$\gamma$ discussed in previous section, suggesting that [Fe\,{\sc ii}] and H$_2$ line emission have similar origin in the central region of NGC\,6240.

\section{Conclusions}

We used near-IR K-band integral field spectroscopy of the inner 1.41$\times$2.35 kpc$^{2}$ of the LIRG NGC\,6240 at a spatial resolution of 90~pc and  velocity resolution of $\sim$40\,\kms, obtained with the NIFS at the Gemini North telescope, to map the molecular and ionized gas emission line flux distributions and kinematics, as well as the stellar kinematics. The main conclusions of this work are:
\begin{itemize}
\item The stellar velocity field show a velocity amplitude of $\sim$200\,\kms\ for the northern nucleus and $\sim$400~\kms\ for the southern nucleus. The velocity dispersion map shows values ranging from 150 to 350~\kms, with the highest values observed at the region between both nuclei. 

\item The H$_2\lambda2.12\,\mu$m and \br\ line emission is observed in the whole field of view, while the coronal line emission (traced by the [Ca\,{\sc viii}]$\lambda2.32\,\mu$m) is observed mainly at the locations of the nuclei. 

\item The gas kinematics is complex and the emission lines show more than one component at locations between the nuclei and surrounding the northern nucleus, while at the southern nucleus and at other regions one component is enough to reproduce each emission line.  


\item Thermal processes play an important role on the origin of H$_2$ line emission at most locations. For locations next to both nuclei, the heating of the gas by X-rays from the AGN may represent the main excitation mechanism, while in locations between both nuclei, the H$_2$ excitation is dominated by shocks. Fluorescent excitation may contribute with a fraction of the H$_2$ emission at 1$^{\prime\prime}$ west from the northern nucleus.

\item Shocks and X-rays from the AGN may also be the origin of the \feii\ emission, as the \feii/\br\ and \hm/\br\ line ratios show a similar trend. The coronal line emission is originated from gas ionized by the radiation of the AGN, as it is observed only at locations next to the nuclei.

\end{itemize}

\section*{Acknowledgments}
 We thank the referee for his/her thorough review, comments and
suggestions, which helped us to significantly improve this paper.
Based on observations obtained at the Gemini Observatory, 
which is operated by the Association of Universities for Research in Astronomy, Inc., under a cooperative agreement with the 
NSF on behalf of the Gemini partnership: the National Science Foundation (United States), the Science and Technology 
Facilities Council (United Kingdom), the National Research Council (Canada), CONICYT (Chile), the Australian Research 
Council (Australia), Minist\'erio da Ci\^encia e Tecnologia (Brazil) and south-eastCYT (Argentina).  
This research has made use of the NASA/IPAC Extragalactic Database (NED) which is operated by the Jet
 Propulsion Laboratory, California Institute of  Technology, under contract with the National Aeronautics and Space Administration.
The authors acknowledges support from FAPERGS (project N0. 2366-2551/14-0) and CNPq (project N0. 470090/2013-8 and 302683/2013-5).

\end{document}